\documentclass[page-classic]{epl2} 

\title{Small anisotropy of the lower critical field and the $s_\pm$-wave two-gap feature in single crystal LiFeAs}
\shorttitle{Small anisotropy of the lower critical field and $s_\pm$-wave two-gap feature in LiFeAs}

\author{Yoo Jang Song\inst{1} \and Jin Soo Ghim\inst{1} \and Jae Hyun Yoon\inst{1} \and Kyu Joon Lee\inst{2} \and Myung Hwa Jung\inst{2} \and Hyo-Seok Ji\inst{3} \and Ji Hoon Shim\inst{3} \and Yunkyu Bang\inst{4}
\and Yong Seung Kwon\inst{1}\thanks{E-mail:\email{yskwon@skku.ac.kr}}}
\shortauthor{Yoo Jang Song \etal}

\institute{
  \inst{1} Department of Physics, Sungkyunkwan University, Suwon 440-746, Republic of Korea\\
  \inst{2} Initiative Center for Superconductivity, Department of Physics,Sogang University, Seoul 121-741, Republic of Korea\\
  \inst{3} Department of Chemistry, Pohang University of Science and Technology, Pohang 790-784, Republic of Korea\\
  \inst{4} Department of Physics, Chonnam National University, Kwangju 500-757, Republic of Korea
}

\abstract{
The in- and out-of-plane lower critical fields and magnetic penetration depths of LiFeAs were examined. The anisotropy ratio $\Gamma_{H_{c1}}(0)$ was smaller than the expected theoretical value, and increased slightly with increasing temperature from 0.6$T_c$ to $T_c$. The small degree of anisotropy was numerically confirmed by considering electron correlation effect. The temperature dependence of the penetration depths followed a power law ($\sim$$T^n$) below 0.3$T_c$, with $n$$>$3.5 for both $\lambda_{ab}$ and $\lambda_c$. Based on theoretical studies of iron-based superconductors, these results suggest that the superconductivity of LiFeAs can be represented by an extended $s_\pm$-wave due to the weak impurity scattering effect. And the magnitudes of the two gaps were also evaluated by fitting the superfluid density for both the in- and out-of-plane to the two-gap model. The estimated values for the two gaps are consistent with the results from angle resolved photoemission spectroscopy and specific heat experiments.}

\begin{document}

\maketitle

\section{Introduction}
Since the discovery of iron-based superconductors, considerable effort has been devoted to identifying a higher transition temperature ($T_c$) and examining the mechanism for the superconducting (SC) behavior. Determining the symmetry of the superconducting order-parameter (OP) is an important first step towards unveiling the SC mechanism in iron-based materials. The magnetic penetration depth $\lambda(T)$ is one of the most useful parameters for probing the symmetry of the OP. The SC gap structure and symmetry for 1111- and 122-type superconductors have been studied by estimating the magnetic penetration depth using tunneling diode resonators (TDR)\cite{Martin_prl09,Gordon_prb10,Malone}, microwave methods\cite{Shibauchi,Hashimoto1,Hashimoto2}, and scanning SQUID susceptometry\cite{Hicks,Luan}.
Previous studies on the magnetic penetration depth in SmFeAsO$_{1.8}$F$_{0.2}$\cite{Malone}, PrFeAsO$_{1-y}$\cite{Hashimoto1}, and Ba$_{1-x}$K$_x$Fe$_2$As$_2$\cite{Hashimoto2} reported an exponential temperature dependence, supporting the existence of a $s$-wave-type SC gap. On the other hand, the magnetic penetration depth for LaFePO\cite{Hicks} exhibited a near linear temperature dependence, supporting a nodal gap. Recent reports on Fe-based superconductors, however, revealed a power law behavior with an exponent of 2$\leq$$n$$\leq$2.5 at low temperatures\cite{Gordon_prb09,Gordon_prb10}. Nevertheless, symmetric properties of SC gaps, such as extended $s$-wave\cite{Luan}, $s_\pm$-wave\cite{Sknepnek,Machida,Vovontsov,Liu}, nodal $d$-wave\cite{Hicks,Fletcher09}, and  point nodal gap~\cite{Gordon_prb09}, have been suggested. Recent theoretical advances\cite{Bang,Dolgov}that consider an impurity scattering effect in the Eliashberg equation have provided convincing evidence that the $s_\pm$-wave SC state is the most promising candidate for the true pairing state of the Fe-pnictide superconductors.
For anisotropy of Fe based superconductors, experimental measurements have revealed very small anisotropy ratios compared to those of cuprate, although both superconductors have the same layered structure\cite{Ni,Jia}. Furthermore, some theoretical studies have suggested the importance of electron correlations\cite{Haule} and their effects on the unconventional superconductivity\cite{Mazin,Dai} of Fe pnictide.

In this letter, we show that both magnetic penetration depths, $\lambda_{ab}$ for $H$$\parallel$$c$ and $\lambda_c$ for $H$$\parallel$$ab$, follow a power law ($\sim$$T^n$) with $n$$>$3.5 below 0.3$T_c$. These results suggest that the superconductivity of LiFeAs can be represented by an extended $s$$_\pm$-wave model due to a weak impurity scattering effect. In addition, the anisotropy ratio of lower critical field is found to be smaller than the expected theoretical value. According to first principles calculations for LiFeAs compounds, the anisotropy ratios of $\Gamma_\lambda(0)$ and $\Gamma_\rho(0)$, estimated using the magnetic penetration depth and electrical resistivity are approximately 3 and 9, respectively\cite{Nakamura}. We also show that the anisotropy ratio can be examined by considering the electron correlation effect on the band structure using a local density approximation (LDA) and dynamical mean field theory (DMFT) approach. Finally, we report that LiFeAs possess two SC gaps determined through fittings to superfluid density for the in- and out-of-plane directions.

\section{Experiment}
The single crystal growth of LiFeAs with $T_c^{zero}$=17.5 K is detailed Ref.\cite{Song}. The single crystal used in this study was obtained from another batch of the same ingot referred to in Ref.\cite{Song}. The magnetic field dependence of the magnetization $M(H)$ was scanned from -70 to 70 kOe every 0.5 K from 2 to 7 K, and every 1 K from 7 K to $T_c$, for both $H$$\parallel$$ab$ and $H$$\parallel$$c$ using a vibrating sample magnetometer (VSM SQUID, Quantum Design).

\section{Results and Discussions}

\begin{figure}[tbp]
\includegraphics[width=1 \linewidth]{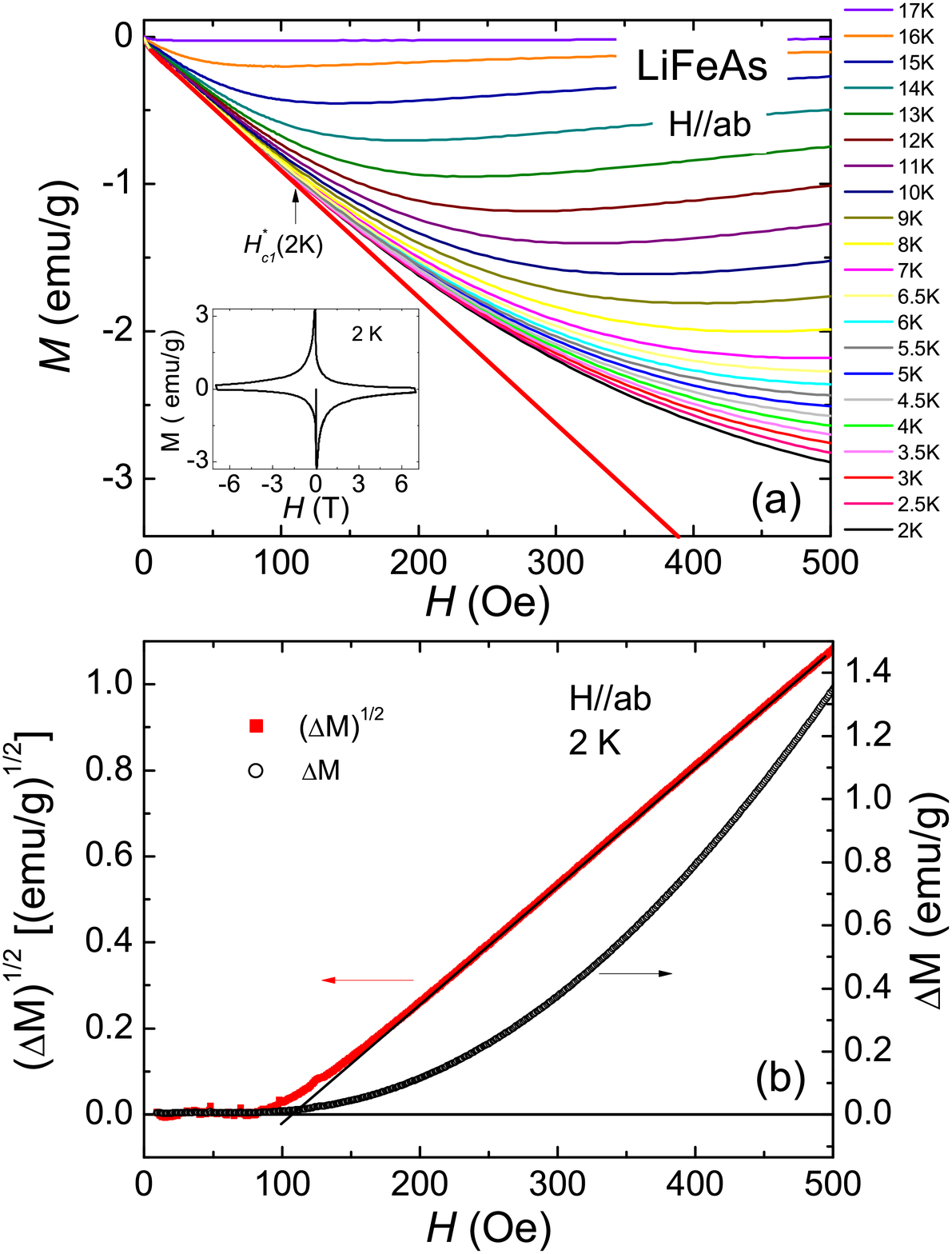}
\caption{\label{figure:epsart}(Color online) (a) Initial region of the magnetization curve $M(H)$ of single-crystal LiFeAs  for $H$$\parallel$$ab$ at various temperatures. Inset: $M$-$H$ hysteresis loop at 2 K
(b) $\Delta M$ as a function of $H$ and $(\Delta M)^{1/2}$ as a function of $H$ at 2 K, where $\Delta M$ is the difference between the Meissner line and observed magnetization.}
\end{figure}


The inset in Fig.1(a) shows the field dependence of the local magnetization measured at 2 K after zero field cooling. The shape of the magnetization loop is almost symmetric about the horizontal axis, which suggests that the hysteresis arises mainly from bulk flux pinning rather than from the surface barrier. Fig.1(a) shows magnetization curves $M(H)$ for $H$$\parallel$$ab$ in the low field range below $T_c$. The $M(H)$ data for $H$$\parallel$$c$ shows  similar behavior(not shown here). The value of $H_{c1}^*$ was obtained by evaluating the deviation of magnetic field from the Meissner line in the initial slope of the $M(H)$ curve. As shown in Fig.1(a), the magnetization very gradually deviates from the Meissner line, making the accurate location of $H_{c1}^*$ difficult to locate. To avoid any ambiguity in locating $H_{c1}^*$ from these broadened $M$-$H$ curves, the following expression is employed \\

   $B_{eq}(H)$= $(2A)^{1/2}$${(H+4\pi M)^{1/2}}$   ~~ ($H_{c1}$$<$$H$$<$$H^*$).   (1)\\

Here, $B_{eq}(H)$ is the equilibrium flux density, $A$ is a constant, $H^*$ is the field defined as $B_{eq}(H^*)$=$A$, and $H$+4$\pi M$($\equiv$ 4$\pi \Delta M$) is the amount corresponding to the deviation of the observed magnetization from the Meissner line. Eq.(1) indicates that, in principle, the plot of $(\Delta M)^{1/2}$ versus $H$ should give the equilibrium $B$-$H$ curve. Therefore, the lower critical field is given by the threshold field of this plot\cite{Bean}. On the other hand, Eq.(1) assumes that the current density is field-independent. A previous study reported that the critical density of LiFeAs is independent of the magnetic field in low field regions below 700 Oe\cite{Song}, and this provides a reliable way of determining the value of $H_{c1}^*$.
Fig.1(b) shows plots of $\Delta M$ versus $H$ and $(\Delta M)^{1/2}$ versus $H$. The threshold of the $(\Delta M)^{1/2}$ plot may be slightly smeared due to the demagnetization field at the sharp corners of the specimen. Therefore, $H_{c1}^*$ can be determined by extrapolating the linear part of the $(\Delta M)^{1/2}$ versus $H$ curve to intersect the horizontal axis\cite{Naito}. The $H_{c1}^*$ value is different from the actual lower critical field $H_{c1}$ due to geometric effects. The well-adapted relationship between $H_{c1}^*$ and $H_{c1}$ proposed by Brandt \cite{Brandt} was adopted in this case: $H_{c1}$=$H_{c1}^*$/tanh($\sqrt{0.36b/a}$), where $a$ and $b$ are the width and thickness of the slab-like superconductor, respectively. Here, $a$=2.8 mm and $b$=1.3 mm for $H$$\parallel$$c$, whereas $a$=1.3 mm and $b$=2.8 mm for $H$$\parallel$$ab$. These values then yield $H_{c1}^{\parallel c}$=2.58$H_{c1}^{\parallel c*}$ and $H_{c1}^{\parallel ab}$=1.41$H_{c1}^{\parallel ab*}$, respectively. The lower critical fields parallel and perpendicular to the plane obtained for each temperature are plotted in Fig.2 as a function of temperature.

\begin{figure}[tbp]
\includegraphics[width=1 \linewidth]{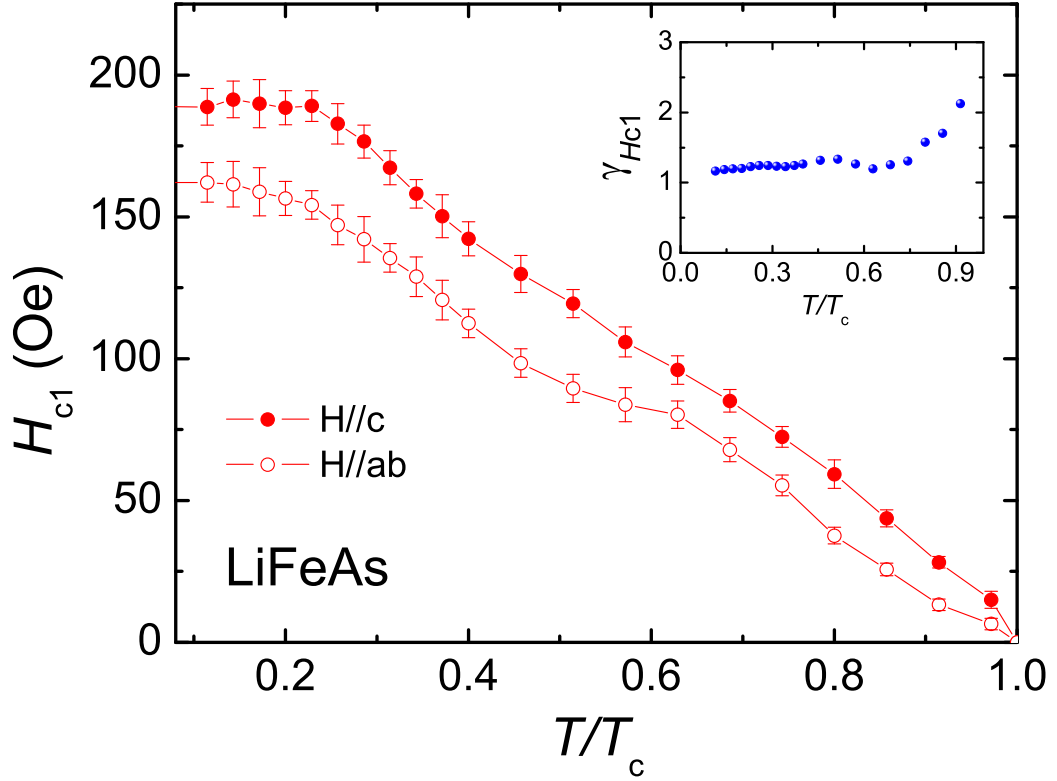}
\caption{\label{fig:epsart}(Color online). Extracted $H_{c1}$ of LiFeAs as a function of $T$ for $H$$\parallel$$ab$ and $H$$\parallel$$c$. The bars show the uncertainty of estimated by the deviating point of the $M$-$H$ curve from the Meissner line in Fig. 1(a). The inset shows the temperature dependence of the anisotropy of $H_{c1}$: $\Gamma_{H_{c1}}$=$H_{c1}^{\parallel c}/H_{c1}^{\parallel ab}$.
}
\end{figure}
$H_{c1}(0)$ was evaluated by extrapolating $H_{c1}(T)$ in the low temperature region. As shown in Fig. 2, the flat behavior below 0.2$T_c$ and the existence of an inflection point near 0.5$T_c$ suggest that LiFeAs is a superconductor that has a full gap without nodes and a two-gap structure. The inset in Fig. 2 shows that the anisotropy of $H_{c1}$, $\Gamma_{H_{c1}}$=$H_{c1}^{\parallel c}/H_{c1}^{\parallel ab}$, has a small value (between 1.2 and 2), and that $\Gamma_{H_{c1}}$ increases slightly with increasing temperature above 0.6$T_c$. The temperature dependence of $\Gamma_{H_{c1}}$ was similar to that observed for MgB$_2$\cite{Lyard} and PrFeAsO$_{1-y}$\cite{Shibauchi,Hashimoto1}, indicating multiband superconductivity. Such behavior may be due to the existence of Fermi surfaces (FSs) of different sizes and anisotropies\cite{Singh}.

\begin{figure}[tbp]
\includegraphics[width=1 \linewidth]{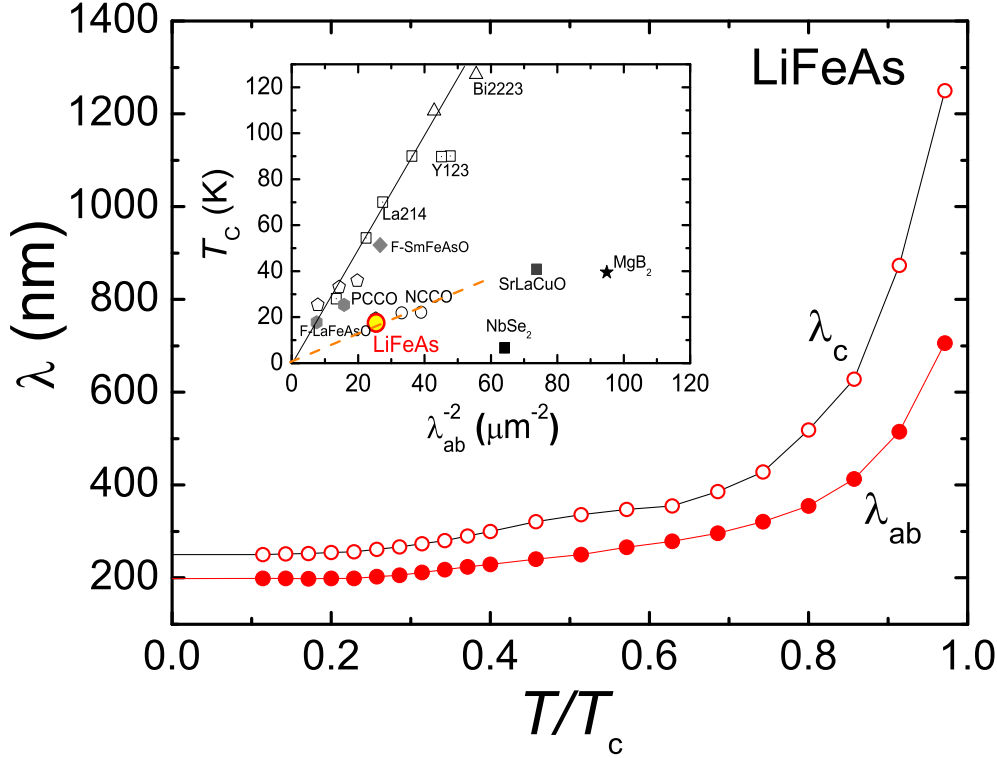}
\caption{\label{fig:epsart} (Color online) Temperature dependence of the magnetic penetration depths $\lambda_{ab}$ and $\lambda_c$ of LiFeAs. Inset: Uemura plot of the superconducting transition temperature $T_c$ versus $\lambda_{ab}^{-2}$ for LiFeAs.}
\end{figure}
The following London formulae were used to estimate the penetration depth : \\
$H_{c1}^{\parallel c}$=$(\Phi_0/4\pi\lambda_{ab}^2)$$\ln\kappa_c$, and $H_{c1}^{\parallel ab}$=$(\Phi_0/4\pi\lambda_{ab}\lambda_c)\ln\kappa_{ab}$, where $\Phi_0$ is the flux quantum and the Ginzburg-Landau parameter $\kappa_c$=$\lambda_{ab}/\xi_{ab}$, $\kappa_{ab}$=$\sqrt{\lambda_{ab}\lambda_c/\xi_{ab}\xi_c}$.
Figure 3 shows the temperature dependence of the magnetic penetration depths, $\lambda_{ab}$ for $H$$\parallel$$c$ and $\lambda_c$ for $H$$\parallel$$ab$, of LiFeAs. The depths at T=0 were obtained by extrapolating $\lambda(T)$ in the low temperature region, and were calculated to be $\lambda_{ab}(0)$$\approx$198.4 nm and $\lambda_c(0)$$\approx$250 nm. As shown in the inset of Fig.3, although these are small errors associated with these values, the relationship between $T_c$ and $\lambda_{ab}^{-2}(0)$=25.4 $\mu$m$^{-2}$ is well represented by the electron-doped Uemura plot (dotted red line), and is also consistent with the results for the transverse-field muon-spin rotation (TF-$\mu$SR) of two polycrystalline LiFeAs compounds~\cite{Pratt}.
The anisotropy of the penetration depth $\Gamma_\lambda$=$\lambda_c/\lambda_{ab}$ exhibited small values of 1.26 at 0 K and 1.8 near $T_c$;similar to the behavior seen for $\Gamma_{H_{c1}}$.

\begin{table}
\caption{Calculated renormalization factors for each Fe 3$d$ orbital obtained using the LAD+DMFT approach.}
\label{tab.1}
\begin{center}
\begin{tabular}{|c|c|c|c|c|c|}
\hline
  Orbital&$d_{z^2}$&$d_{x^2-y^2}$&$d_{xz,yz}$&$d_{xy}$ \\
\hline
  1/$z$=$m^*/m$ & 2.09 & 2.08 & 2.89 & 3.86 \\
\hline
\end{tabular}
\end{center}
\end{table}

To confirm the small anisotropy of the LiFeAs compound, the electron correlation effect on the band structure was examined using a LDA+DMFT approach. The LDA calculations were performed using the full-potential linearized augmented plane-wave method in the Wien2k code\cite{Blaha}, and DMFT was implemented based on the LDA Hamiltonian method. The impurity problem within the DMFT self-consistency equation was solved using the continuous time quantum Monte Carlo calculation. Parameters that give excellent agreement with the ARPES experiments\cite{Borisenko} were used: $U$=5.0 eV for the Coulomb interaction, and $J$=0.54 eV for the Hund's rule coupling constant. In addition, the specific heat coefficient $\gamma$ was calculated to be 29.1 mJ/mol$\cdot$K$^2$. This is in good agreement with the experimental value of 35 mJ/mol$\cdot$K$^2$, obtained from $C/T$$=$$\gamma$$+$$\beta T^2$ in the measurement of the low temperature specific heat for the our sample(data not shown). Owing to the electron correlation effect, the ARPES of LiFeAs showed strong band renormalization: a factor of three larger compared to that of the LDA bands. Because each Fe 3$d$ orbital has a different hybridization strength, each orbital also has a unique renormalization factor $m^*/m$ (where $m^*$ is the effective mass). Table 1 lists the calculated renormalization factors of each Fe 3$d$ orbital. The smallest renormalization factors are those of the $d_{z^2}$ and $d_{x^2-y^2}$ orbitals, whereas the $d_{xy}$ orbital has the largest renormalization factor. Owing to the difference in the renormalization factors, the band renormalization can be varied according to the orbital character in momentum space. Because the electrical anisotropy is strongly dependent on the band dispersion anisotropy, different band renormalizations can result in a large modification of the electrical anisotropy compared to that obtained from LDA calculations. Indeed, the LDA+DMFT calculation showed that $\rho_c/\rho_{ab}$= 2.19, which is significantly smaller than the result of $\rho_c/\rho_{ab}$=9 obtained by LDA\cite{Nakamura}. The result from the LDA+DMFT approach for LiFeAs corresponds with the electrical anisotropy ratio estimated from the experimental results:  $\Gamma_\rho$$\sim$3.3\cite{Song}.
\begin{figure}[tbp]
\includegraphics[width=1 \linewidth]{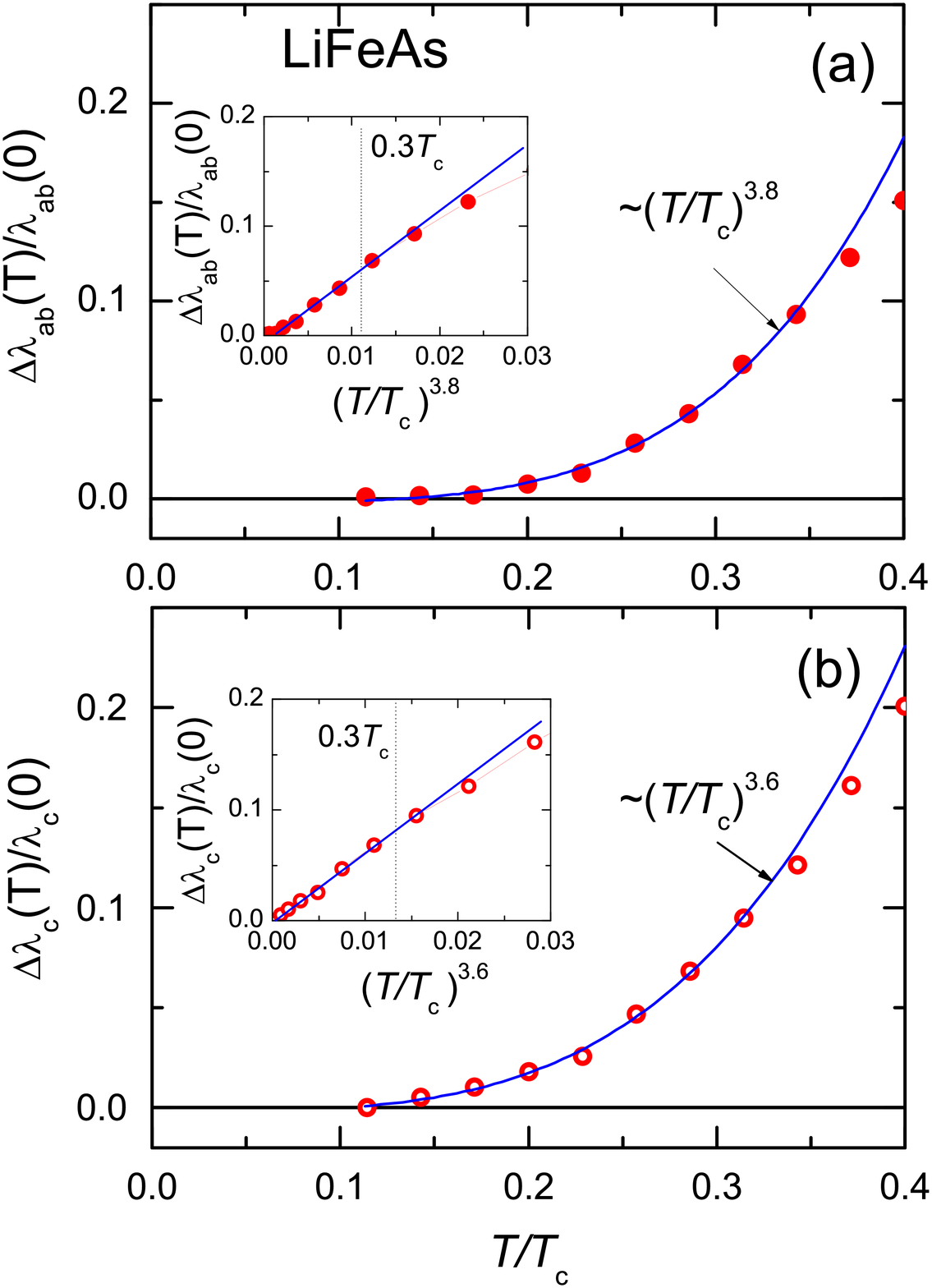}
\caption{\label{fig:epsart} (Color online) Temperature dependence of $\Delta\lambda(T)/\Delta\lambda(0)$ for the penetration depths $\lambda_{ab}$ and $\lambda_c$. The solid line(blue) is $\Delta\lambda\sim(T/T_c)^n$
.}
\end{figure}

Fig.4 shows values of $\Delta\lambda(T)$ estimated from the superfluid densities $\rho^s(T)$=$H_{c1}(T)/H_{c1}(0)$
=$\lambda^2(0)/\lambda^2(T)$ of the in- and out-of-plane directions; note that $\Delta\lambda(T)$=$\lambda(T)$-$\lambda(0)$. In the low-temperature region below 0.3$T_c$, the power law $\Delta\lambda(T)$$\sim$$(T/T_c)^n$ (solid blue line) provides a suitable fit for both $\Delta\lambda_{ab}(T)$ and $\Delta\lambda_c(T)$, with $n$=3.8$\pm$0.1 for $H$$\parallel$$c$ and $n$=3.6$\pm$0.1 for $H$$\parallel$$ab$, respectively. In previous reports on doped Fe-based superconductors, the penetration depth behavior could be modeled using a power law with an exponent value of 2$\leq$$n$$\leq$2.5 in the low temperature region\cite{Gordon_prb09,Gordon_prb10}. Many authors of theoretical articles have argued, based on the power-law behavior of the penetration depth at low temperatures, that most pnictides are $s$$_\pm$-wave superconductors due to the impurity scattering effect\cite{Bang,Dolgov}. According to Ref.\cite{Dolgov}, the low temperature power-law behavior of the undoped LiFeAs superconductor, with an exponent of $n$$>$3.5, suggests that the superconductor could be represented by an extended $s$$_\pm$-wave model due to the weak impurity scattering effect. These results also correspond closely to the recent results of NMR and NQR experiments for polycrystalline Li$_x$FeAs compounds\cite{Li}.\\
\begin{figure}[tbp]
\includegraphics[width=1 \linewidth]{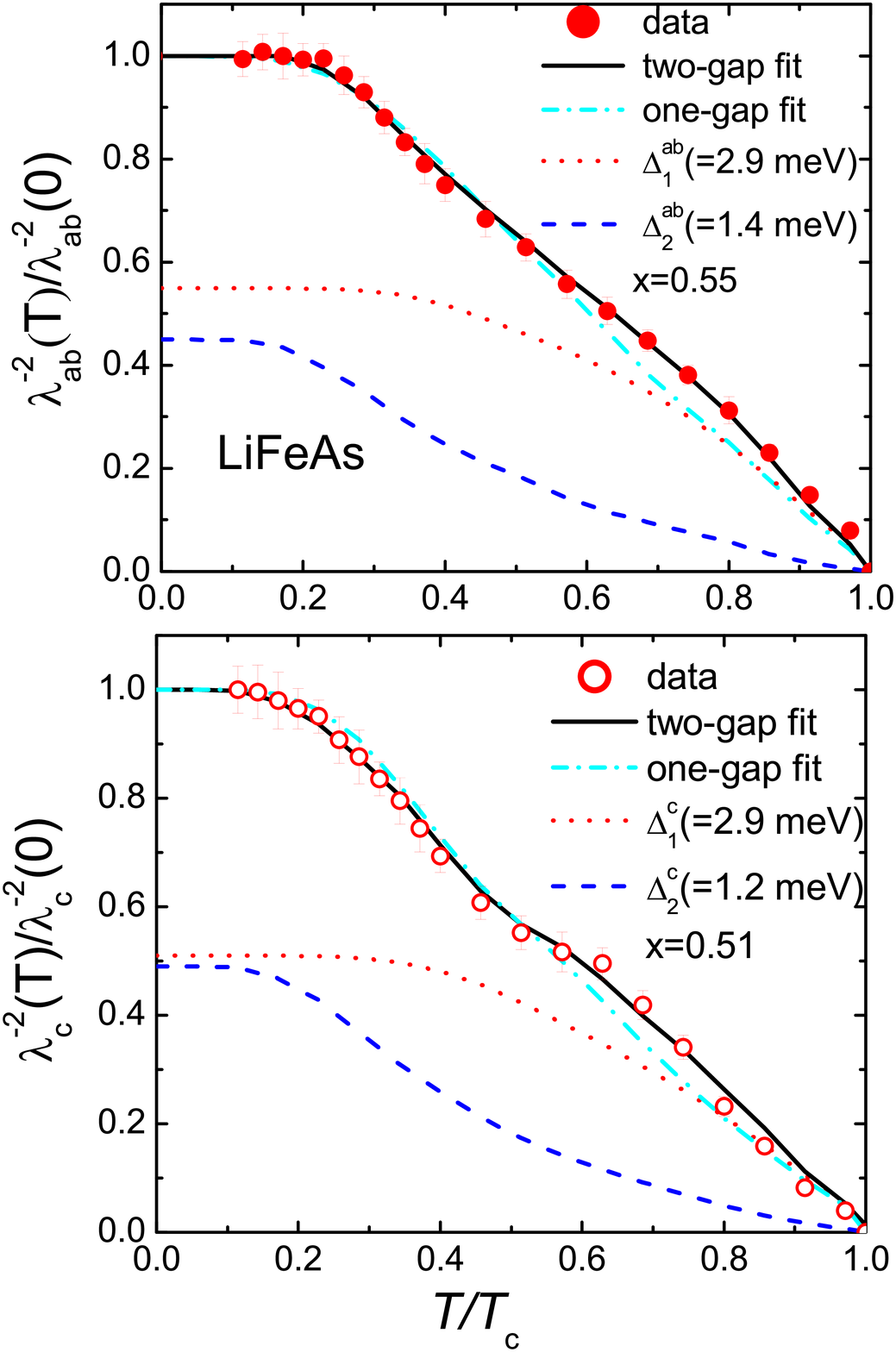}
\caption{\label{fig:epsart} (Color online) Temperature dependence of the superfluid density $\lambda^2(0)$/$\lambda^2(T)$ for the in-plane and out-of-plane directions in LiFeAs. The solid lines (black) are the best fit for the s$_\pm$-wave two-gap model(Eq.(2)).}
\end{figure}

Recently, the $s$-wave two-gap feature has been examined using angle resolved photoemission spectroscopy (ARPES)\cite{Borisenko}, specific heat\cite{Wei}, and lower critical field\cite{Sasmal} experiments.
In this study, the following expressions were used to examine the two-gap feature of LiFeAs:\\

$\tilde{\rho_s}(T)$=$x\tilde{\rho_s}(T,\bigtriangleup_1(0))$+$(1-x)\tilde{\rho_s}(T,\bigtriangleup_2(0))$   $~~$    (2)\\

$\tilde{\rho_s}(T)$=1-2$\int_{\bigtriangleup(T)}^\infty( -\frac{\partial f(E)}{\partial E})\frac{E}{\sqrt{E^2-\bigtriangleup(T)^2}}dE$   $~~~~~~~$(3)\\

Here, $x$ (0$\leq$$x$$\leq$1) is a weighting factor that indicates the contribution of the large gap, and $f(E)$=[1+exp$(E/k_BT)$]$^{-1}$ is the Fermi function. The temperature dependence of the gap $\bigtriangleup(T)$ can be expressed as $\bigtriangleup(T)$=$\bigtriangleup(0)$$\tanh$[1.82[1.018($T_c/T$-1)]$^{0.51}]$\cite{Carrington}.

As shown in Fig.5(a) and (b), the results of the fitting with two gaps (solid black line) were more consistent with the experimental data than those with one gap (dash-dotted blue line). The following parameters were used in the model: $\bigtriangleup_1^{ab}(0)$=2.9$\pm$0.2 meV, $\bigtriangleup_2^{ab}(0)$=1.4$\pm$0.1 meV, $x$=0.55 for $H$$\parallel$$c$, and $\bigtriangleup_1^c(0)$=2.9$\pm$0.2 meV, $\bigtriangleup_2^c(0)$=1.2$\pm$0.1 meV, $x$=0.51 for $H$$\parallel$$ab$.  These results were in good agreement with those obtained using ARPES\cite{Borisenko,Stockert}, specific heat\cite{Wei}, lower critical field\cite{Sasmal}, microwave surface impedance measurements\cite{IMAI}, and TDR experiments\cite{Kim}. Sasmal \emph{et al}\cite{Sasmal} reported the lower critical field, anisotropy and two-gap features of LiFeAs using a method similar to that reported in Ref.\cite{Ren}. In the present study, however, the temperature dependence of $\Gamma_{H_{c1}}$ due to the multiband effect, SC gap symmetry and isotropic properties of the gaps are novel, and the small measured anisotropy ratio was also confirmed using the LDA+DMFT approach by considering the electron correlation effect. On the other hand, these results are in contrast with the single gap analysis reported by Inosov \emph{et al}\cite{Inosov}.

\section{Summary}
In summary, this study has confirmed the small anisotropy ratio of the lower critical fields and penetration depth in LiFeAs by both magnetic measurements and numerical calculations using the LDA+DMFT approach and considering the electron correlation effect. The low temperature power-law behavior ($n$$>$3.5) of the penetration depth suggests that the LiFeAs superconductor can be represented using an extended $s_\pm$-wave model due to the weak impurity scattering effect. In addition, the sizes of the two SC gaps in LiFeAs for the in- and out-of-plane were obtained. The results were consistent with those reported previously.

\acknowledgments
This study was supported by the Basic Science Research Program (2010-0007487, 2010-0006484), the Mid-career Researcher Program (No. R01-2008-000-20586-0) and the Nuclear R$\&$D Programs (2006-2002165 and 2009-0078025) through the National Research Foundation of Korea (NRF) funded by the Ministry of Education, Science and Technology.

\end{document}